\documentclass[12pt]{article}

\usepackage{color}

\usepackage{graphicx}
\usepackage{float}
\usepackage{amsmath}
\usepackage{amsthm}
\usepackage{amssymb}
\usepackage{amscd}
\usepackage{amsfonts}
\usepackage{makeidx}
\usepackage{enumerate}
\newtheorem{teo}{Theorem}[section]
    \newtheorem{lem}[teo]{Lemma}
    \newtheorem{prop}[teo]{Proposition}
    \newtheorem{coro}[teo]{Corollary}
    \newtheorem{defn}[teo]{Definition}
    \newtheorem{obs}[teo]{Remark}

\newcommand{\re}{\ref}

\newcommand{\ci}{\cite}

    \newtheorem*{dem}{\textsc{Proof}}
    \newcommand{\bdem}{\begin{dem}}
    \newcommand{\edem}{\end{dem}}
     \newcommand{\be}{\begin{equation}}
    \newcommand{\ee}{\end{equation}}
     \newcommand{\ba}{\begin{array}}
    \newcommand{\ea}{\end{array}}
\newcommand{\beqn}{\begin{eqnarray}}
    \newcommand{\eeqn}{\end{eqnarray}}
    \newcommand{\bl}{\begin{lem}}
    \newcommand{\el}{\end{lem}}
    \newcommand{\bp}{\begin{prop}}
    \newcommand{\ep}{\end{prop}}
\newcommand{\ds}{\displaystyle}
     
     \newcommand{\al}{\alpha}

     \newcommand{\om}{\omega}
     
    \newcommand{\R}{\mathbb{R}}
    \newcommand{\C}{\mathbb{C}}
     
     \newcommand{\no}{\noindent}
     \newcommand{\Res}{{\rm Res}}

\def\Re {{\rm Re\, }}
 \def\Im {{\rm Im\,}}
\def\Res {{\rm Res\,}}

\begin{document}
\title{Time-dependent approach
to the  uniqueness of the Sommerfeld
 solution
of the diffraction problem by a half-plane}
\author{
\large{A.  Merzon*  $^1$},\\
\large{ P.  Zhevandrov$^2$},\\
\large{J.E. De la Paz M\'endez$^3$}\\
\large{T.J. Villalba Vega$^4$}\\
{\small{\it $^1$ Instituto de F\'\i sica y  Matem\'aticas, Universidad Michoac{a}na}},\\[-2mm]
{\small  Morelia, Michoac\'{a}n, M\'{e}xico},\\[-1mm]
anatolimx@gmail.com
\\[-1mm]
{\small{\it $^2$ Facultad de  Ciencias F\'\i sico-Matem\'aticas, Universidad Michoac{a}na}},\\[-2mm]
{\small  Morelia, Michoac\'{a}n, M\'{e}xico}\\[-2mm]
{\small{\it $^3$ Facultad de Matem\'aticas II, Universidad Aut\'onoma de Guerrero}},\\[-2mm]
{\small{Cd. Altamirano, Guerrero, M\'exico}}\\[-2mm]
{\small{\it $^4$ Universidad Aut\'onoma de Guerrero, Campus Taxco}},\\[-2mm]
{\small{Taxco, Guerrero, M\'exico}},\\[-2mm]}


\maketitle


\begin{abstract}
We consider the Sommerfeld problem of diffraction  by an opaque half-plane with a real wavenumber
 interpreting it as the limiting case, as time tends to infinity, of the corresponding time-dependent diffraction problem. We prove that the Sommerfeld formula for the solution is the limiting amplitude of the solution of this time-dependent problem  which belongs to a certain functional class and is unique in it. For the proof of  uniqueness of solution to the time-dependent problem we reduce it, after the Fourier-Laplace transform in $t$, to a stationary diffraction problem with a complex wavenumber. This permits us to use the proof of  uniqueness in the Sobolev space $H^1$. Thus we avoid imposing the radiation  and regularity conditions on the edge from the beginning and instead obtain it in a natural way.
\end{abstract}


\section{Introduction}
\setcounter{equation}{0}

The main goal of this paper is to prove the uniqueness of solution to the Sommerfeld half-plane problem \cite{Som1896}- \cite{NZG} with a real wavenumber, proceeding from the uniqueness of the corresponding time-dependent problem in a certain functional class. The existence and uniqueness of solutions to this problem was considered in many papers, for example in \cite{P.S}-\cite{ST}. However, in our opinion, the problem of uniqueness is still not solved in a satisfactory form from the point of view of  boundary value problems.\;The fact is that this problem is  a homogeneous boundary value problem which admits various nontrivial solutions. Usually the ``correct" solutions are chosen by physical reasoning \cite{Som1896}-\cite{P.S}, for example using the Sommerfeld radiation conditions and regularity conditions on the edge.\\
The question is: from the mathematical point of view, where from the radiation and the regularity conditions arise? Our goal is to show that they arise automatically from the \textit{nonstationary} problem as conditions for the limiting amplitude for the latter. Of course, the limiting amplitude principle (LAP) under suitable conditions is very well-know for the diffraction by smooth obstacles, see e.g. \ci{Eidus1969}-\ci{Vain},  but we are unaware of its rigorous proof in the case of diffraction by a half-plane.

\no The literature devoted to diffraction by wedges including the Sommerfeld problem is enormous (see e.g. reviews in \cite{g4} and \cite{KMEMV}), and we will only indicate some papers where the uniqueness is treated. In paper \cite{P.S} a uniqueness theorem was proven for the Helmholtz equation $(\Delta+1)u=0$ in two-dimensional regions $D$ of the semiplane type. These regions can have a finite number of bounded obstacles with singularities on their boundaries. In particular, the uniqueness of solution $u$ to the Sommerfeld problem was proven by means of the decomposition of the solution into the sum $u=g+h$, where $g$ describes the geometrical optics incoming and reflected waves and $h$ satisfies the Sommerfeld radiation condition (clearly, $u$ should also satisfy the regularity conditions al the edge).

\no In paper \cite{ST} exact conditions were found for the uniqueness in the case of complex wave number. The problem was considered in Sobolev spaces for a wide class of generalized incident waves, and for $DD$ and $NN$ boundary conditions.
In paper \cite{HA} the same problem was considered also for the complex wave number and for $DN$-boundary conditions. In both papers the Wiener-Hopf method has been used.
Finally in \cite {g4} along with the proof of existence, the uniqueness of solution of BVP for the Helmholtz equation with complex wavenumber in arbitrary angle was proven in the Sobolev space. Note that this result is fundamental to our construction in this paper.

\no Time-dependent scattering by wedges was considered in many papers although their number is not so large as the number of papers devoted to the stationary scattering by wedges. We indicate here the following papers: \cite{SS}-\cite{Rot}. The detailed description of these papers is given in \cite{ktime}.\\
In \cite{KMEMV}, \cite{ktime}-\cite{la},  the  diffraction   by a wedge of magnitude $\phi$ (which can be a half-plane in the case $\phi=0$ as in \cite{KMEMV}) with a real wavenumber was considered as a stationary problem which is the  ``limiting case" of a \textit{nonstationary} one. More precisely, we seeked the solutions of the classical diffraction problems as \textit{limiting amplitudes} of solutions to corresponding nonstationary problems, which are unique in some appropriate functional class. We also, as in \cite{P.S},  decomposed the solution of nonstationary problem separating  a  ``bad" incident wave, so that the other part of solution belongs to a certain appropriate functional class. Thus we avoided the {\it apriori} use of the radiation and regularity conditions and instead obtained them in a natural way.
In papers  \cite{ktime}- \cite{mesqNN} we considered the time-dependent scattering with DD, DN and NN boundary conditions and proved the uniqueness of solution in an appropriate functional class. But these results were obtained only for $\phi\neq0$ because in the proof of uniqueness we used the Method of Complex Characteristics \cite{K-1973}-\cite{kmz} which ``works" only for $\phi\neq0$.

\no For $\phi=0$ we need to use other methods, namely, the reduction of the uniqueness problem for the stationary diffraction to the uniqueness problem for the corresponding time-dependent diffraction, which in turn is reduced to the proof of uniqueness of solution of stationary problem but with a {\it complex wavenumber}.\\
Note that in \cite{la} we proved the LAP for $\phi\neq0$ and for the $DD$-boundary conditions. Similar results for the $NN$ and $DN$ b.c were obtained in \cite{mm}-\cite{mesqNN}. A generalization of these results to the case of generalized incident wave (cf. \cite{ST}) were given in \cite{ktime}. This approach (stationary diffraction as the limit of time-dependent one) permits us to justify all the known classical explicit formulas 
\cite{SS}-\cite{KB} and to prove their coincidence with the explicit formulas given in \cite{ktime}, \cite{kmm}, \cite {mesqNN}.  In other words, all the classical known formulas are the limiting amplitudes of solutions to nonstationary problems as $t\to\infty$. For the Sommerfeld problem, this was proven in \cite{KMEMV}, except for the proof of the uniqueness of the solution to the nonstationary problem in an appropriate class. This paper makes up for this omission.

\begin{figure}[htbp]
\centering
\includegraphics[scale=0.3]{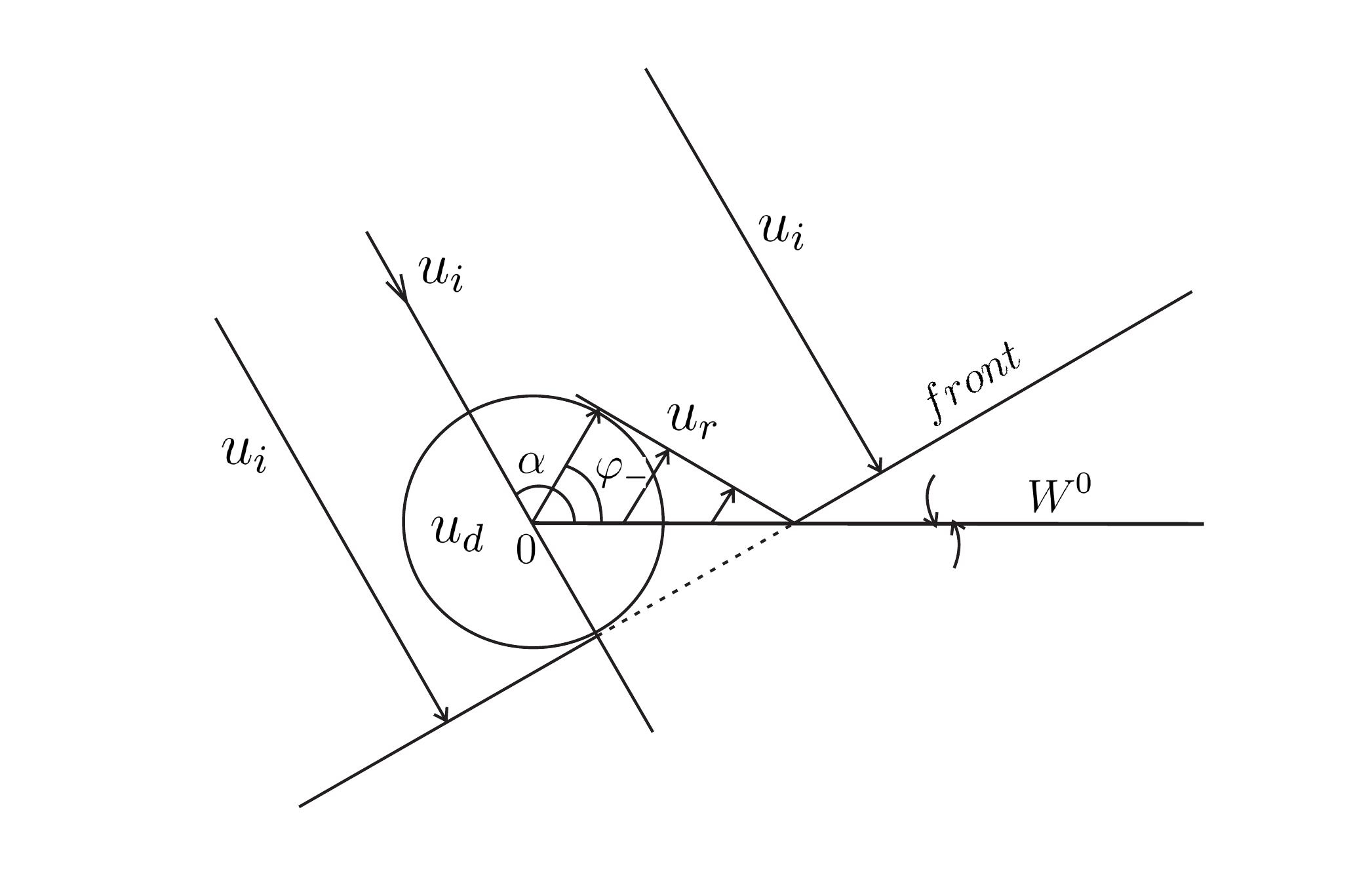}
\caption{Time-dependent diffraction by a half-plane}\label{Figdif}
\end{figure}

\no Our plan is as follows. The nonstationary diffraction problem is reduced by means of the Fourier-Laplace transform with respect to time $t$ to a stationary one with a complex wave number. For this problem the uniqueness theorems can be proven more easily in Sobolev spaces and do not use the radiation and regularity conditions. Then we prove
that the Fourier-Laplace transforms of solutions to nonstationary diffraction half-plane problem, whose amplitude tends to the Sommerfeld solution, also belong to a Sobolev space for a rather wide class of incident waves. This permits us to  reduce the problem to the case of \cite{g4}.\\
Let us pass to the problem setting. We consider the
two-dimensional time-dependent scattering of a plane wave by the half-plane $W^{0}:=\lbrace(x_1,x_2)\in\R^2:x_2=0, x_1>0\rbrace$.
The nonstationary incident plane wave in the absence of obstacles reads
\be\label{u_i}
u_i(x,t) = e^{-i\om_{0} (t - {\bf n}\cdot x)} f( t-{\bf n}\cdot x),\quad x\in\mathbb{R}^2,\quad t\in\mathbb{R},
\ee
where
\begin{equation}\label{n}
\omega_0>0,~ {\bf n}=(n_1,n_2)=(\cos(\pi+\al),\sin(\pi+\al)),
\end{equation}
and $f$ is ``a profile function'', such that $f\in L^1_{loc}(\R)$, and
\begin{equation}\label{f}
f(s)=0,\quad s< 0,\quad\sup(1+|s|)^p|f(s)|<\infty,\quad{\rm for~some}~p\in\R,
\quad \lim\limits_{s\to+\infty}f(s)=1.
\end{equation}
\begin{obs}\label{ru_i}
The incident wave $u_i$ given by (\ref{u_i}) belongs to a wide class which includes all the functions (\ref{f}), in particular non-periodic functions (when $\omega_0=0$). Obviously, these functions satisfy the D'Alembert equation $\square u_{i}(x,t)=0$ in the distribution sense.
\end{obs}
\no For definiteness, we assume that
\begin{equation}\label{alpha}
\frac{\pi}{2}<\alpha<\pi.
\end{equation}
In this case the front of the incident wave $u_i$ reaches the half-plane $W^0$ for the first time at the moment $t=0$ and at this moment  the reflected wave $u_r(x,t)$  is born (see Fig. \ref{Figdif}). Thus
\begin{equation*}
u_{r}(x,t)\equiv0,\quad t<0.
\end{equation*}
Note, that for $t\to\infty$ the amplitude of $u_i$ is exactly equal to the Sommerfeld incident wave \cite{som54} by (\ref{f}), see also (\ref{0.5}) below.

\no The time-dependent scattering with the Dirichlet boundary conditions  is described by the mixed problem
\begin{equation}\label{mp}
\left\{     \begin{array}{rcl}
              \square u(x,t):=(\partial_t^2-\Delta)   u(x,t) = 0, &  & x\in Q \\\\
              u(x,t) = 0, &  & x\in\partial Q
            \end{array}
\right|    \ t\in\mathbb{R},
\end{equation}
where $Q:=\R^2\setminus W^{0}$. The ``initial condition'' reads
\begin{equation}\label{ic}
u(x,t)=u_{i}(x,t),\quad x\in Q,\quad t<0,
\end{equation}
where $u_{i}$ is the incident plane wave (\ref{u_i}).\\
Introduce the nonstationary ``scattered'' wave $u_{s}$ as the difference between $u$ and $u_{i}$,
\begin{equation}\label{u_s'}
u_{s}(x,t):=u(x,t)-u_{i}(x,t),\quad x\in Q,\quad t\in \mathbb{R}.
\end{equation}

\vspace{0.5cm}

\no Since $\square u_{i}(x,t) =0,~ (x,t)\in Q\times \mathbb{R}$, we get from (\ref{ic}), (\ref{mp}) that
\begin{eqnarray}\label{D1}
\square u_{s}(x,t) &=& 0, \quad (x,t)\in Q\times \mathbb{R}
                                \\  \nonumber\\ \label{icus}
u_{s}(x,t)&=& 0 ,\quad x\in Q,\quad t<0
\\   \nonumber\\ \label{bc}
u_{s}(x,t)&=& -u_{i}(x,t),\quad x\in W^{0},\quad t>0.
\end{eqnarray}
Denote
\begin{equation}
\label{varphi_+}
\varphi_{\pm}:=\pi\pm\alpha.
\end{equation}
Everywhere below we assume that
\begin{equation}\label{x=}
x_1=r\cos\varphi, \quad x_2=r\sin\varphi,\quad 0\leq\varphi<2\pi.
\end{equation}
Let us define the nostationary incident wave in the presence of the obstacle $W^0$ which is the opaque screen,
\begin{equation}\label{u_i^0}
 u_{i}^{0}(\rho,\varphi,t):=\left\{\ba{rccc}
& u_{i}(\rho,\varphi,t),& \quad  0<\varphi<\varphi_{+}&\\\\
 & 0, &  \quad \varphi_{+}<\varphi<2\pi.&
\ea
\right.
\end{equation}
\begin{obs}\label{obs}
The function $u_{s}$ has no physical sense, since $u_{i}\neq u_{i}^{0}$. The wave $u_{s}$ coincides with the scattered wave $u_{s}^{0}:=u-u_{i}^{0}$ in the zone $\lbrace (\rho,\varphi): 0<\varphi<\varphi_{+}\rbrace$, but in the zone $\varphi_{+}<\varphi\leq2\pi$ we have $u_{s}^{0}=u_{s}+u_i$.
\end{obs}

\no The goal of the paper is to prove that the Sommerfeld solution of half-plane diffraction problem is the limiting amplitude of a solution to time-dependent problem (\ref{mp}), (\ref{ic}) (with any $f$ satisfying (\ref{f})) and this solution is unique in an appropriate functional class. The paper is organized as follows.\\
In Section\;2 we recall the Sommerfeld solution. In Section\;3 we reduce the time-dependent diffraction problem to a stationary one and define a functional class of solutions. In Section\;4 we give an explicit formula for the solution of the time-dependent problem and prove that the Sommerfeld solution is its limiting amplitude. In Section\;5 we prove that the solution belongs to the corresponding functional class. Finally, in Section\;6 we prove the uniqueness.

 \section{Sommerfeld's diffraction}
 \setcounter{equation}{0}

Let us recall the Sommerfeld solution \cite{som54}, \cite{NZG}. The stationary incident wave (rather, the incident wave amplitude)  is
\begin{equation}\label{0.5}
\mathcal{A}_i^{0}(\rho,\varphi)=e^{-i\omega_0\rho\cos(\varphi-\alpha)},\quad\varphi\in[0,\varphi_+].
\end{equation}
We denote this incident wave as $\mathcal{A}_i^0$ since it is the limiting amplitude of the nonstationary incident wave $u_i^0$ given by (\ref{u_i^0}):
$\mathcal{A}_i^0(\rho,\varphi)= \lim\limits_{t\to\infty}e^{i\omega_0t}u_i^0(x,t)$,
in view of formula (\ref{u_i}), see Remark \ref{obs}.
The Sommerfeld half-plane diffraction problem can be formulated as follows: to find a function $\mathcal{A}(x), x\in\overline{Q}$ such that
\begin{equation}\label{BV1}
\left\{\ba{rcl}
(\Delta+\omega^2)\mathcal{A}(x)&=&0,\quad x\in Q\\\\
\mathcal{A}(x)\Big\vert_{W^0}&=&0
\ea
\right.
\end{equation}
\begin{equation}\label{mathcal{A}}
\mathcal{A}(x)=\mathcal{A}_i^0(x)+\mathcal{A}_r(x)+\mathcal{A}_d(x),\quad x\in Q,
\end{equation}
where $\mathcal{A}_r(x)$ is the reflected wave,
\begin{equation}\label{mathcal{A}_r(x)}
\mathcal{A}_r(x)=-e^{-i\omega_0\rho\cos(\varphi+\alpha)},\quad \varphi\in(0,\varphi_{-}),
\end{equation}
and $\mathcal{A}_d(x)$ is the wave diffracted by the edge,
\begin{equation}\label{Cad}
\mathcal{A}_d(x)\to 0,\quad |x|\to\infty.
\end{equation}
A. Sommerfeld \cite{som54} found the solution of this problem, given by the formula
\begin{equation*}
\mathcal{A}(\rho,\varphi)=\frac{1}{4\pi}\int\limits_{\mathcal{C}}\zeta(\gamma,\varphi)e^{-i\omega\rho\cos\gamma}d\gamma,\quad \rho\geq0,\quad \varphi\in[0,2\pi],
\end{equation*}
where
\begin{equation}\label{0.11}
\zeta(\gamma,\varphi):=\Big( 1-e^{i(-\gamma+\varphi-\alpha)/2}\Big)^{-1}-\Big( 1-e^{i(-\gamma+\varphi+\alpha)/2}\Big)^{-1},\quad \gamma\in\C
\end{equation}
and $\mathcal{C}$ is the Sommerfeld contour (see (\cite[formula (1.1) and Fig. 3]{KMEMV}).\\
In the rest of the paper we prove that this solution is a limiting amplitude of a solution of time-dependent problem (\ref{mp}) and is unique in an appropriate functional class.\\
The Sommerfeld diffraction problem can also be considered for NN and DN half-plane. The corresponding formulas for solution can be founded in \cite{ktime}.\\
Sommerfeld has obtained his solution using an original method of solutions of the Helmholtz equation on a Riemann surface. Note that a similar approach was used for the wedge diffraction of the rational angle \cite{Sp} where well-posedness in suitable Sobolev space was proved.

\section{Reduction to a ``stationary" problem. Fourier-Laplace transform. }
\setcounter{equation}{0}

Let $\hat{h}(\omega), \omega\in\C^+$ denote the Fourier-Laplace transform $\mathcal{F}_{t\to\omega}$ of $h(t)$,
\begin{equation}\label{F}
\hat{h}(\omega)=\mathcal{F}_{t\to\omega}[h(t)]=\int_{0}^{\infty} e^{i\omega t} h(t)\;dt,\quad h\in L_{1}(\overline{\mathbb{R}}_{+});
\end{equation}
$\mathcal{F}_{t\to\omega}$ is extended by continuity to $S'(\overline{\mathbb{R}}_{+})$. Assuming that for any $x\in Q$, $u_s(\bullet,t)\in S'(\overline{\R^+})$ (see (\ref{icus})), we apply this transform to system (\ref{D1})-(\ref{bc}), and obtain
\begin{equation}\label{SP}
    \left\{\ba{rcl}
            (\Delta + \om^2)\hat{u}_{s}(x,\omega) &=&0, \quad x\in Q,   \\\\
                     \hat{u}_{s}(x,\omega) &=& -\hat{u}_{i}(x,\omega),\quad x\in W^{0}
        \ea
\right| \quad \omega\in\mathbb{C}^{+}.
\end{equation}
Let us calculate $\hat{u}_{i}(x,\omega)$. Changing the variable $t-{\bf n}\cdot x=\tau$, and using the fact that ${\rm supp} f\subset \overline{\mathbb{R}^{+}}$ we obtain from (\ref{u_i}) and (\ref{n}) that
\begin{equation}\label{hat{u}_i(x,om)}
\hat{u}_{i}(x,\omega)=e^{i\omega {\bf n}\cdot x} \hat{f}(\omega-\omega_0).
\end{equation}
Hence,
\begin{equation*}
\hat{u}_{i}(x_1,0,\omega)=e^{i\omega n_1 x_1}\hat{f}(\omega-\omega_0),\quad x_1>0,
\end{equation*}
and the boundary condition in (\ref{SP}) is $\hat{u}_{s}(x_1,0,\omega)=-g(\omega) e^{i\omega n_1 x_1}$. Therefore we come to the following family of BVP depending on $\omega\in \mathbb{C}^{+}$: to find $\hat{u}_{s}(x,\omega)$ such that
\begin{equation}\label{BVP}
    \left\{
    \ba{rclcc}
            (\Delta + \om^2)\hat{u}_{s}(x,\omega) &=&0,&\quad  x\in Q \\\\
                     \hat{u}_{s}(x_1,0,\omega) &=& -g(\omega) e^{i\omega n_1 x_1},&\quad x_1>0.
        \ea
\right.
\end{equation}
We are going to prove the existence and uniqueness of solution to problem (\ref{mp}), (\ref{ic}) such that $u_s$ given by (\ref{u_s'}) belongs to the space $\mathcal{M}$, which is defined as follows:
\begin{defn}\label{d1}
$\mathcal{M}$ is the space of functions $u(x,t)\in S'(\overline{Q}\times\overline{\R^{+}})$ such that its Fourier-Laplace transform $\hat{u}(x,\omega)$ is a holomorphic function on $\omega\in\C^{+}$ with values in $C^{2}(Q)$ and
\begin{equation}\label{em}
\hat{u}(\cdot, \cdot, \omega)\in H^{1}(Q)
\end{equation}
for any $\omega\in\C^{+}$.

\end{defn}

\begin{obs}\label{3.2}
Note that $u_i(x, t)\Big\vert_{\overline{Q}\times\overline{\R^+}}\notin\mathcal{M}$ since $\big|e^{i\omega {\bf n}\cdot x}\big|=  e^{\omega_2\rho\cos(\varphi-\alpha)}$ and for $0<\alpha-\varphi<\pi/2$, it  grows exponentially as $\rho\to\infty$, and hence does not satisfy (\ref{em}); because of this we use system (\ref{D1})-(\ref{bc}) instead of (\ref{mp}) (they are equivalent by (\ref{ic})) since (\ref{D1})-(\ref{bc}) involves only the values of $u_i$ on the boundary and the latter possess the Fourier-Laplace transform which do not grow exponentially.
\end{obs}

\begin{obs}\label{3.2'}
Since for $u_s\in H^1(Q)$ the Dirichlet and Neumann data exist in the trace sense and in the distributional sense respectively (see e.g; \cite{g4}), problem (\ref{BVP}) is well-posed. Hence, problem (\ref{D1})-(\ref{bc}) is well-posed too.
\end{obs}

\section{Connection between the nonstationary diffraction problem (\ref{mp}), (\ref{ic}) and the Sommerfeld half-plane problem.}

\setcounter{equation}{0}

In paper \cite{KMEMV} we solved problem (\ref{mp}), (\ref{ic}). Let us recall the corresponding construction. First we define the nonstationary reflected wave \cite[ formula (26)]{KMEMV}
\begin{equation}\label{u_{r}}
    u_{r}(x,t)=\left\{\ba{rcl}
            &-e^{-i\omega_0(t-\overline{{\bf n}}\cdot x)} f(t-\overline{{\bf n}}\cdot x)&, \quad \varphi\in(0,\varphi_{-})   \\\\
                     &0& ,\quad \varphi\in(\varphi_{-},2\pi)
        \ea
\right| \quad t\geq0,
\end{equation}
where $\overline{{\bf n}}:=(n_1,-n_2)=(-\cos\alpha,\sin\alpha)$ (see Fig. \re{Figdif}).\\
Note that its limiting amplitude coincides with (\ref{mathcal{A}_r(x)}) similarly to the incident wave.

\no Second, we define the nonstationary diffracted wave (cf. \cite[ formula (31) for $\phi=0$]{KMEMV}).

\medskip
\no Let
\begin{equation}\label{CZ}
\mathcal{Z}(\beta,\varphi):=Z(\beta+2\pi i-i\varphi),
\end{equation}
and
\begin{equation}\label{u_d}
u_{d}(\rho,\varphi,t)=\frac{i}{8\pi}\int_{\mathbb{R}}\mathcal{Z}(\beta,\varphi)  F(t-\rho\cosh\beta)\;d\beta,
\end{equation}
where $\varphi\in(0,2\pi)$, $\varphi\neq\varphi_{\pm}$; $t\geq0$,
\begin{equation}\label{F(s)}
F(s)=f(s) e^{-i\omega_{0} s},
\end{equation}
\begin{equation}\label{Z^{0}}
Z(z)= -U\Big(-\frac{i\pi}{2}+z\Big)+U\Big(-\frac{5i\pi}{2}+z\Big),
\end{equation}
\begin{equation}\label{H^{0}}
U(\zeta)=\coth\Big(q(\zeta-i\frac{\pi}{2}+i\alpha)\Big)-\coth\Big(q(\zeta-i\frac{\pi}{2}-i\alpha)\Big),\quad q=\frac{1}{4}
\end{equation}
for the Dirichlet boundary conditions. Below in Lemma \ref{Z^0(z_)} we give the necessary properties of the function $\mathcal{Z}$, from which the convergence of  integral (\ref{u_d}) follows. Obviously, the condition ${\rm supp~} F\subset [0,\infty)$ (see (3.1)) implies that
${\rm supp~}u_d (\cdot,\cdot,t)\subset [0,+\infty)$.
\begin{obs}\label{math J}
The function $U(\gamma+\varphi)$ essentially coincides with the Sommerfeld kernel (\ref{0.11}). This is for a reason. In paper \cite{kmm} it was proven that the solution to the corresponding time-dependent diffraction problem by an arbitrary angle $\phi\in(0,\pi]$ belonging to a certain class similar to $\mathcal{M}$ necessarily has the form of the Sommerfeld type integral with the Sommerfeld type kernel.
\end{obs}
\no Finally, we proved \cite[Th. 3.2, Th 4.1]{KMEMV} the following
\begin{teo}\label{loc}
i) For $f\in L_{loc}^{1}(\mathbb{R})$ the function
\begin{equation}\label{u_{ird}}
u(\rho,\varphi,t):=u_{i}^{0}(\rho,\varphi,t)+u_{r}(\rho,\varphi,t)+u_{d}(\rho,\varphi,t),\quad \varphi\neq\varphi_{\pm}
\end{equation}
belongs to $L_{loc}^{1}(Q\times\mathbb{R}^{+})$. It is continuous up to $\partial Q\times\mathbb{R}$ and satisfies the boundary conditions and initial conditions (\ref{mp}), (\ref{ic}).
The D'Alembert equation in (\ref{mp}) holds in the sense of distributions.\\
ii) The LAP holds for Sommerfeld's diffraction by a half-plane:
$\lim\limits_{t\to\infty}e^{i\omega_0t}u(\rho,\varphi,t)=\mathcal{A}(\rho,\varphi),\quad\varphi\neq\varphi_\pm$.
\end{teo}
\no Since the main object of our consideration will be the ``scattered" wave $u_s(x,t)$ given by (\ref{u_s'}), we clarify the connection between $u_s$ and the Sommerfeld solution $\mathcal{A}$.
\begin{coro}\label{CUS}
Define $\mathcal{A}_i(x)= e^{-i\omega_0\rho\cos(\varphi+\alpha)}$, which is the limiting amplitude of $u_{i}(x,t)$ given by (\ref{u_i}). The limiting amplitude of $u_s(x,t)$ is the function
\begin{equation}\label{mathcal{A}_s(x)}
\mathcal{A}_s(x)=\mathcal{A}(x)-\mathcal{A}_i(x),
\end{equation}
i.e. $\lim_{t\to\infty} e^{i\omega_0 t} u_s(x,t)=\mathcal{A}_s(x)$.
\end{coro}
\no {\bf Proof.}  The statement follows from (\ref{u_s'}). ~~$\blacksquare$
\begin{obs}
The function $\mathcal{A}_s$ is the amplitude of the scattered non-stationary wave $u_s(x,t)$ and $\mathcal{A}_s$ satisfies the following nonhomogeneous B.V.P.
\begin{equation}\label{BV2}
    \left\{\ba{rcl}
            &(\Delta+\omega_0) \mathcal{A}_s(x)=0&, \quad x\in Q   \\\\
                     &\mathcal{A}_s\Big\vert_{W^0}=-\mathcal{A}_i(x)&.
        \ea
\right.
\end{equation}
This BVP (as well as (\ref{BV1})) is ill-posed since the homogeneous problem admits many solutions (i.e.,  the solution is nonunique).
\end{obs}
\begin{obs}\label{R4.4}
$\mathcal{A}_s$ can be decomposed similarly to (\ref{mathcal{A}}). Namely, by (\ref{mathcal{A}_s(x)}), (\ref{mathcal{A}}) we have
\begin{equation}\label{bv3}
\mathcal{A}_s=\mathcal{A}_i^0+\mathcal{A}_r(x)+\mathcal{A}_d(x)-\mathcal{A}_i(x)=\mathcal{A}_r(x)+\mathcal{A}_d(x)-\mathcal{A}_i^{1}(x),
\end{equation}
where $\mathcal{A}_i^{1}(x)=\mathcal{A}_i(x)-\mathcal{A}_i^{0}(x)$.
Obviously,  problems (\ref{BV2}), (\ref{bv3}) and (\ref{BV1}), (\ref{mathcal{A}}) with the condition (\ref{Cad}) are equivalent, but the first problem is more convenient as we will see later.
\end{obs}

\section{Solution of the ``stationary"  problem.}
\setcounter{equation}{0}

In this section we will obtain explicit formula for the solution of (\ref{BVP}) and prove that it belongs to $H^1(Q)$ for all $\omega\in\C^+$.\\
Let $\mathcal{Z}(\beta,\varphi)$ be given by (\ref{CZ}). First, we will need the Fourier-Laplace transform of the reflected and diffracted waves (\ref{u_{r}}), (\ref{u_d}).

\begin{lem}\label{le3}
The Fourier-Laplace transform of $u_r$ and $u_d$ are
\begin{equation}\label{^u_r}
   \hat{u}_{r}(x,\omega)=\left\{\ba{rcl}
             -\hat{f}(\omega-\omega_0)e^{-i\omega\rho\cos(\varphi+\alpha)},& \quad\varphi\in(0,\varphi_{-})   \\\\
                     0,& \quad\varphi\in(\varphi_{-},2\pi),
        \ea
\right.
\end{equation}

\medskip
\begin{equation}\label{^u_d}
\hat{u}_{d}(\rho,\varphi,\omega)=\frac{i}{8\pi}\hat{f}(\omega-\omega_0)\int_{\mathbb{R}} \mathcal{Z}(\beta,\varphi)\;e^{i\omega\rho\cosh\beta}\;d\beta,\quad \omega\in\mathbb{C}^{+},\quad \varphi\neq\varphi_{\pm}.
\end{equation}
\end{lem}
\no{\bf Proof.} From (\ref{u_{r}}) we have
\begin{equation*}
    \hat{u}_{r}(x,\omega)=\left\{\ba{rcl}
            & -\mathcal{F}_{t\to\omega}\Big[e^{-i\omega_{0}(t-\overline{{\bf n}}\cdot x)} f(t-\overline{{\bf n}}\cdot x)\Big]&, \quad \varphi\in(0,\varphi_{-})   \\\\
                     &0& ,\quad \varphi\in(\varphi_{-},2\pi).
        \ea
\right.
\end{equation*}
Further $-\mathcal{F}_{t\to\omega}\Big[e^{-i\omega_{0}(t-\overline{{\bf n}}\cdot x)} f(t-\overline{{\bf n}}\cdot x)\Big]=-e^{i\omega_{0}(\overline{{\bf n}}\cdot x)}\int_{0}^{\infty} e^{i(\omega-\omega_{0})t} f(t-\overline{{\bf n}}\cdot x)\;dt$.\\
Changing the variable $t-\overline{{\bf n}}\cdot x=\tau$,
we obtain $\hat{u}_{r}(x,\omega)=-e^{i\omega\;\overline{{\bf n}}\cdot x} \ds\int_{-\overline{{\bf n}}\cdot x}^{\infty} e^{i(\omega-\omega_{0})\tau} f(\tau)\;d\tau,\quad\varphi\in(0,\varphi_{-})$.
Moreover, by (\ref{u_{r}}) $-\overline{{\bf n}}\cdot x=\rho\cos(\varphi-\alpha)\leq c<0,\quad \varphi\in(0,\varphi_{-})$, since $\pi/2<\alpha<\varphi+\alpha<\pi$ by (\ref{alpha}) and (\ref{varphi_+}). Hence, we obtain (\ref{^u_r}),
since ${\rm supp} f\subset\overline{\mathbb{R}^{+}}$. The second formula in (\ref{^u_r}) follows from definition (\ref{u_{r}}) of $u_r$.\\
\no Let us prove (\ref{^u_d}). Everywhere bellow we put  $\omega=\omega_1+i\omega_2$, $\omega_{1,2}\in\R$, $\omega_2>0$, for $\omega\in\C^+$.  By Lemma \ref{Z^0(z_)}(i), (\ref{f}) and (\ref{F(s)}) we have
\begin{equation*}
\Big| e^{i\omega t}\mathcal{Z}(\beta,\varphi)F(t-\rho\cosh\beta)\Big|\leq C e^{-\omega_2 t} e^{-\beta/2} (1+t)^{-p},\quad\rho<0,~\varphi\neq\varphi_\pm,~\beta\in\R.
\end{equation*}
Hence, by the Fubini Theorem there exists the Fourier-Laplace transform of $u_d(\cdot,\cdot,t)$ and
\begin{equation}\label{hat{u}_d}
\hat{u}_{d}(\rho,\varphi,\omega)=\frac{i}{8\pi}\int_{\mathbb{R}}\mathcal{Z}(\beta,\varphi)\mathcal{F}_{t\to\omega}\Big[ F(t-\rho\cosh\beta)\Big]\;d\beta,\quad \varphi\neq\varphi_\pm.
\end{equation}
We have
\begin{equation*}
G(\rho,\beta,\omega):=\mathcal{F}_{t\to\omega}\Big[ F(t-\rho\cosh\beta)\Big]=
\int_{0}^{\infty} e^{i\omega t} F(t-\rho\cosh\beta)\Big]\;dt,\quad \omega\in\mathbb{C}^{+}.
\end{equation*}
Making the change of the variable $\tau=t-\rho\cosh \beta$ in the last integral and using the fact that ${\rm supp}\;F \subset [0,\infty)$ and $\hat{F}(\omega)=\hat{f}(\omega-\omega_0)$ by (\ref{F(s)}), we get $\ds G(\rho,\beta,\omega)=e^{i\omega\rho\cosh\beta}\hat{f}(\omega-\omega_0)$. Substituting this expression into (\ref{hat{u}_d}) we obtain (\ref{^u_d}). Lemma \ref{le3} is proven. ~~~$\blacksquare$

\subsection{Estimates for $\hat{u}_{r}, \partial_{\rho}\hat{u}_{r}, \partial_{\varphi}\hat{u}_{r}$.}

\begin{lem}\label{leu_r}
For any $\omega\in\C$, there exist $C(\omega), c(\omega)>0$, such that both functions $\hat{u}_{r}$ and $\partial_\rho\hat{u}_{r}$ admit the same estimate

\begin{equation}\label{e^u_r}
    \left.\ba{rcl}
    \Big|\hat{u}_r(\rho,\varphi,\omega)\Big|&\leq &C(\omega)e^{-c(\omega)\rho}\\
    \\
    \Big|\partial_\rho\hat{u}_r(\rho,\varphi,\omega)\Big|&\leq& C(\omega)e^{-c(\omega)}
          \ea
\right|
\quad \rho>0,\quad\varphi\in (0,2\pi), \quad\varphi\neq\varphi_\pm.
\end{equation}
and $\partial_\varphi\hat{u}_{r}(\rho,\varphi,\omega)$ admits the estimate
\begin{equation}\label{d+u_r}
|\partial_\varphi\hat{u}_{r}(\rho,\varphi,\omega)|\leq C(\omega)\rho\; e^{-c(\omega)\rho},\quad \rho>0.
\end{equation}
\end{lem}
\noindent{\textbf{Proof.}}  By (\ref{alpha}) there exits $c(\omega)>0$ such that
$
\ds\left|e^{-i\omega\rho\cos(\varphi+\alpha)}\right|=e^{\omega_2\rho\cos(\varphi+\alpha)}\leq e^{-c(\omega)\rho},\quad 0<\varphi<\varphi_-
$
by (\ref{alpha}). Therefore (\ref{e^u_r}) holds for $\hat{u}_r$.  Hence,  differentiating (\ref{^u_r}) we obtain (\ref{e^u_r}) for $\partial_\rho\hat{u}_r$ and (\ref{d+u_r}) for $\partial_\varphi\hat{u}_r$, for $\varphi\neq\varphi_{-}$.
~~$\blacksquare$

\subsection{Estimates for $\hat{u}_{d}$.}

\begin{prop}\label{pro1}
There exist $C(\omega),c(\omega)>0$ such that,  the function $\hat{u}_d$,  and $\partial_{\rho}\hat{u}_d$, $\partial_{\varphi}\hat{u}_d$ admit the estimate
\begin{equation}\label{5.6'}
    \left.\ba{rcl}
    \Big|\hat{u}_d(\rho,\varphi,\omega)\Big|&\leq& C(\omega)e^{-c(\omega)\rho}\\
    \\
     \Big|\partial_\rho\hat{u}_d(\rho,\varphi,\omega)\Big| &\leq& C(\omega)e^{-c(\omega)\rho}(1+\rho^{-1/2})   \\\\
\Big|\partial_\varphi\hat{u}_d(\rho,\varphi,\omega)\Big| &\leq& C(\omega) e^{-c(\omega)\rho}\rho(1+\rho^{-1/2})
        \ea
\right|
\quad \rho>0,\quad\varphi\in(0,2\pi),\quad \varphi\neq\varphi_\pm.
\end{equation}
\end{prop}
\noindent{\bf{Proof.}} {\bf I.} By (\ref{^u_d}), in order to prove (\ref{5.6'}) for $\hat{u}_d$ it suffices to prove that
\begin{equation}\label{eA}
|A(\rho,\varphi,\omega)|\leq C(\omega)e^{-c(\omega)\rho},
\end{equation}
where
\begin{equation}\label{A}
A(\rho,\varphi,\omega):=\int\limits_{\R} \mathcal{Z}(\beta,\varphi) e^{i\omega\rho\cosh\beta}\;d\beta,\quad \varphi\neq\varphi_\pm.
\end{equation}
Represent $A$ as $A=A_1+A_2$,
where
\begin{equation}\label{A_1}
    \left.\ba{rcl}
    A_1(\rho,\varphi,\omega)&:=&\ds\int\limits_{-1}^{1}\mathcal{Z}(\beta,\varphi) e^{i\omega\rho\cosh\beta}\;d\beta\quad\\
    \\
    A_2(\rho,\varphi,\omega)&:=&\ds\int\limits_{|\beta|\geq1} \mathcal{Z}(\beta,\varphi) e^{i\omega\rho\cosh\beta}\;d\beta\quad
        \ea
\right|
\quad \varphi\in(0,2\pi),\quad \varphi\neq\varphi_\pm.
\end{equation}
The estimate (\ref{eA}) for $A_2$ follows from (\ref{asH^0}) (see Appendix\;I). It remains to prove the same estimate for the function $A_1$. Let
\begin{equation}
\label{varepsilon_{pm}}
\varepsilon_{\pm}:=\varphi_{\pm}-\varphi.
\end{equation}
Representing $A_1$ as
\begin{equation*}
A_1(\rho,\varphi,\omega)=-4\mathcal{K}_0(\rho,w,\varepsilon_+)+4\mathcal{K}_0(\rho,w,\varepsilon_-)+\int\limits_{-1}^{1} \check{\mathcal{Z}}(\beta,\varphi) e^{i\omega\rho\cosh\beta}\;d\beta,
\end{equation*}
where $\mathcal{K}_0$ is defined by (\ref{K}),
we obtain (\ref{eA}) for $A_1$ from Lemma \ref{lcI} i) y (\ref{8.0}).

\medskip
\no {\bf II.} Let us prove (\ref{5.6'}) for $\partial_{\rho}\hat{u}_d$. By (\ref{^u_d}) it suffices to prove that
\begin{equation}\label{eB}
|B(\rho,\varphi,\omega)|\leq C(\omega)e^{-c(\omega)\rho}(1+\rho^{1/2}),\quad\varphi\neq\varphi_{\pm},
\end{equation}
where $
B(\rho,\varphi,\omega):=\ds\int\limits_{\R} \mathcal{Z}(\beta,\varphi)\cosh\beta\;e^{i\omega\rho\cosh\beta}\;d\beta$. Represent $B$ as $B_1+B_2$,
where $B_{1,2}(\rho,\varphi,\omega)$ are defined similarly to (\ref{A_1}),
\begin{equation*}
B_1(\rho,\varphi,\omega):=\ds\int\limits_{-1}^{1} \mathcal{Z}(\beta,\varphi)\cosh\beta e^{i\omega\rho\cosh\beta}\;d\beta, \quad B_2(\rho,\varphi,\omega):=\ds\int\limits_{|\beta|\geq1} \mathcal{Z}(\beta,\varphi)\cosh\beta e^{i\omega\rho\cosh\beta}\;d\beta,
\quad \varphi\neq\varphi_\pm.
\end{equation*}
From (\ref{asH^0}) for $\mathcal{Z}$ we have $|B_2(\rho,\varphi,\omega)|\leq C_1\ds\int\limits_{1}^\infty e^{\beta/2}e^{-\frac{1}{2}\omega_2\rho e^{\beta}}\;d\beta$.\\
Making the change of  the variable $\xi:=\rho e^{\beta}$, we get
\begin{equation*}
\begin{array}{lll}
|B_2(\rho,\varphi,\omega)|
\leq \left\{
\begin{array}{lll }
  C_1(\omega)\rho^{-1/2},    &  \rho\leq1  \\
  \\
  \ds\int\limits_{\rho}^\infty \frac{e^{-\omega_2\xi/2}}{\xi^{1/2}}d\xi,    & \rho\geq1.
\end{array}
\right.
\end{array}
\end{equation*}
Since for $\rho\geq1$, $\ds\int\limits_{\rho}^\infty \frac{e^{-\omega_2\xi/2}}{\xi^{1/2}}d\xi\leq \frac{2}{\omega_2}e^{-\omega_2\rho/2}$,
(\ref{eB}) is proved for $B_2$.\\
\no  It remains to prove estimate (\ref{eB}) for $B_1$. Using (\ref{maathcal{Z}}), (\ref{dk}) we write
\begin{equation*}
B_1(\rho,\varphi,\omega)=-4\mathcal{K}_1(\rho,\omega,\varepsilon_+)+4\mathcal{K}_1(\rho,\omega,\varepsilon_-)+\int\limits_{-1}^{1}\check{\mathcal{Z}}(\beta,\varphi)\cdot\cos\beta\;e^{i\omega\rho\cosh\beta}\;d\beta.
\end{equation*}
Hence, $B_1$ satisfies (\ref{eA}) (and meanwhile (\ref{eB})) by Lemma\;\ref{lcI} (i) and (\ref{8.0}).

\medskip

\no {\bf III.} Let us prove (\ref{5.6'}) for $\partial_{\varphi}\hat{u}_d$. By (\ref{^u_d}) it suffices to prove this estimate for $\partial_{\varphi} A$, where $A$ is given by (\ref{A}). From (\ref{d+}) we have
\begin{equation}\label{dfA}
\partial_{\varphi} A(\rho,\varphi,\omega)=-\omega\rho  A_3(\rho,\varphi,\omega),\quad A_3(\rho,\varphi,\omega)=\int\limits_{\R}\mathcal{Z}(\beta,\varphi)\sinh\beta\;e^{i\omega\rho\cosh\beta}\;d\beta,\quad
\varphi\neq\varphi_\pm.
\end{equation}
\no Similarly to the proof of estimate (\ref{eB}) for $B$, we obtain the same estimate for $A_3$, so, by (\ref{dfA}), the estimate (\ref{5.6'}) follows. Proposition \ref{pro1} is proven.~~~~~$\blacksquare$

\medskip
\no Now define the function
\begin{equation}\label{u_s^0}
u_{s}^{0}(\rho,\varphi,t)=u(\rho,\varphi,t)-u_{i}^{0}(\rho,\varphi,t), \quad \varphi\neq\varphi_+, \quad t>0,
\end{equation}
where $u_i^{0}$ is given by (\ref{u_i^0}).
Then by (\ref{u_{ird}})
\begin{equation}
\label{3.9'}
u_{s}^{0}(\rho,\varphi,t)=u_{r}(\rho,\varphi,t)+u_{d}(\rho,\varphi,t),\quad \varphi\neq\varphi_{\pm},\quad t>0,
\end{equation}
where $u_{r}$ is given by (\ref{u_{r}}) and $u_{d}$ is given by (\ref{u_d}).
\begin{coro}\label{cu^0_s}
Let $\hat{u}_s^0(\rho,\varphi,\omega)$ be the Fourier-Laplace transform of the function $u_s^0(\rho,\varphi,t)$. Then the functions $\hat{u}_s^0$, $\partial_{\rho}\hat{u}_s^0$ and $\partial_{\varphi}\hat{u}_s^0$ satisfy (\ref{5.6'}).
\end{coro}
\no{\bf Proof.} From (\ref{3.9'}) we have
\begin{equation}\label{hatu_s^0}
\hat{u}_{s}^{0}(\rho,\varphi,\omega)=\hat{u}_{r}(\rho,\varphi,\omega)+\hat{u}_{d}(\rho,\varphi,\omega),\quad \varphi\neq\varphi_{\pm},\quad \omega\in\C^{+},
\end{equation}
where $\hat{u}_r$ and $\hat{u}_d$ are defined by (\ref{^u_r}) and (\ref{^u_d}), respectively. Hence the statement follows from Lemma \ref{leu_r} and Proposition \ref{pro1}. ~~$\blacksquare$

\subsection{Estimates for $\hat{u}_{s}(x,\omega)$.}

To estimate $\hat{u}_s$ it is convenient to introduce one more ``part" $u_i^1$ of the nonstationary incident wave $u_i$, namely the difference between $u_i$ and $u_i^0$.

From (\ref{u_s'}) and (\ref{u_s^0}) it follows that
\begin{equation}\label{u_s}
u_{s}(\rho,\varphi,t)=u_{s}^{0}(\rho,\varphi,t)-u_i^1(\rho,\varphi,t),\quad \varphi\neq\varphi_{\pm}
\end{equation}
where $u_i^1(\rho,\varphi,t):=u_{i}(\rho,\varphi,t)-u_{i}^{0}(\rho,\varphi,t)$.
From (\ref{u_i}) and (\ref{u_i^0})
\begin{equation}\label{u_1^1}
u_i^1(\rho,\varphi,t)=\left\{\ba{rcl}
& 0&, \quad  0<\varphi<\varphi_{+}\\\\
 & -u_{i}(\rho,\varphi,t)& , \quad \varphi_{+}<\varphi<2\pi.
\ea
\right.
\end{equation}
By (\ref{hat{u}_i(x,om)})
\begin{equation}\label{dil}
\hat{u}_{i}^{1}(\rho,\varphi,\omega)=\left\{\ba{rcl}
& 0&, \quad  0<\varphi<\varphi_{+}\\\\
 & - \hat{f}(\omega-\omega_0)\;e^{i\omega{\bf n}\cdot x}  & , \quad \varphi_{+}<\varphi<2\pi.
\ea
\right.
\end{equation}
\begin{lem}\label{led}
There exist $ C(\omega), c(\omega)>0$ such that $\hat{u}_i^1$, $\partial_{\rho}\hat{u}_i^1$ satisfy (\ref{e^u_r}) and $\partial_{\varphi}\hat{u}_i^1$  satisfies  (\ref{d+u_r}) for $\varphi\in(0,2\pi),~\varphi\neq\varphi_{\pm}$.
\end{lem}
\no{\bf Proof.}  By (\ref{hat{u}_i(x,om)}) it suffices to prove the statement for $e^{i\omega {\bf n}\cdot x}$ when $\varphi\in(\varphi_{+},2\pi)$. Since $|e^{i\omega {\bf n}\cdot x}|=e^{\omega_2\rho\cos(\varphi-\alpha)},~\varphi\in(\varphi_+,2\pi)$ we have
\begin{equation}
\label{dr}
\partial_\rho e^{\omega_2\rho\cos(\varphi-\alpha)}=\omega_2\cos(\varphi-\alpha)e^{\omega_2\rho\cos(\varphi-\alpha)},\quad \partial_\varphi e^{\omega_2\rho\cos(\varphi-\alpha)}=-\omega_2\rho\sin(\varphi-\alpha)e^{\omega_2\rho\cos(\varphi-\alpha)},
\end{equation}
and for $\varphi\in(\varphi_+,2\pi)$, we have $|e^{\omega_2\rho\cos(\varphi-\alpha)}|\leq e^{-c\omega_2\rho }$, $c>0$,  $\varphi\in(\varphi_+,2\pi)$, because  $\cos(\varphi-\alpha)\leq -c<0$ by (\ref{alpha}). Hence the statement follows from (\ref{dr}). $~~~~~\blacksquare $
\begin{coro}\label{le hat{u}_s}
The functions $\hat{u}_s$, $\partial_{\rho}\hat{u}_{s}$ and $\partial_{\varphi}\hat{u}_s$  satisfy (\ref{5.6'}), for $\varphi\in(0,2\pi),~\varphi\neq\varphi_{\pm}$.
\end{coro}
\no{\bf Proof.} From (\ref{u_s}) it follows that
\begin{equation}\label{hat{u}_s=sigma}
\hat{u}_s(\rho,\varphi,\omega)=\hat{u}_s^{0}(\rho,\varphi,\omega)-\hat{u}_i^1(\rho,\varphi,\omega).
\end{equation}
Thus the statement follows from Corollary \ref{cu^0_s} and Lemma \ref{led}. ~~~~~$\blacksquare$

\medskip
\no It is possible to get rid of the restriction $\varphi\neq\varphi_{\pm}$ in the Corollary \ref{le hat{u}_s}.  Indeed, we have:

\no Let $l_\pm=\{(\rho,\varphi):\rho>0,~\varphi=\varphi_\pm\}$.

\begin{prop}\label{patial_varphi}
The functions $\hat{u}_s(\cdot, \cdot, \omega)$, $\partial_\rho\hat{u}_s(\cdot,\cdot,\omega)$ and $\partial_\varphi\hat{u}_s(\cdot,\cdot,\omega)$ belong to $C^2(Q)$, and satisfy (\ref{5.6'}) in $Q$ (including $l_+\cup l_-$), and
\begin{equation}\label{HE}
(\Delta+\omega^2)\hat{u}_s(\rho, \varphi, \omega)=0,\quad (\rho,\varphi)\in Q,\quad \omega\in\C^+.
\end{equation}
\end{prop}
\no {\bf Proof}. The function $\hat{u}_s(\rho, \varphi, \omega)$ satisfies (\ref{HE}) in $Q\setminus\{l_+\cup l_-\}$.
This follows directly from the explicit formulas (\ref{hat{u}_s=sigma}). In fact, (\ref{hat{u}_s=sigma}) and (\ref{hatu_s^0}) imply
\begin{equation}\label{u^_s}
\hat{u}_s=\hat{u}_r+\hat{u}_d-\hat{u}_i^1.
\end{equation}
The function $\hat{u}_r$ satisfies (\ref{HE}) for $\varphi\neq\varphi_\pm$, $\hat{u}_i^1$ satisfies (\ref{HE}) for $\varphi\neq\varphi_\pm$ by (\ref{u_1^1})  and (\ref{hat{u}_i(x,om)}) and $\hat{u}_d$ satisfies (\ref{HE}) for $\varphi\neq\varphi_\pm$ by (\ref{^u_d}), see Appendix\;II. It remains only to prove that $\hat{u}_s\in C^2(Q)$, because it will be mean that (\ref{5.6'}) holds by Corollary \ref{le hat{u}_s} (and continuity) and (\ref{HE}) holds in $Q$ including $l_{\pm}$.\\
Let us prove this for $\varphi$ close to $\varphi_{-}$. The case of $\varphi$ close to $\varphi_{+}$ is analyzed similarly.\\
Let $h(s)$ be defined in $(\C\setminus\R)\cap B(s^{*})$, where $B(s^{*})$ is a neighborhood of $s^{*}\in\R$. Define the jump of $h$ at the point $s^{*}$ as
$$
\mathcal{J}(h,s^{*}):= \lim_{\varepsilon\to0+} h(s^{*}+i\varepsilon)- \lim_{\varepsilon\to0+}h(s^{*}-i\varepsilon).
$$
We have $\mathcal{J}\Big(\hat{u}_r(\rho,\varphi,\omega),\varphi_-\Big)=\hat{f}(\omega-\omega_0) e^{-i\omega\rho}$ by (\ref{^u_r}). Similarly $\mathcal{J}\Big(\partial_{\varphi}\hat{u}_r(\rho,\varphi,\omega),\varphi_-\Big)=0,\quad \mathcal{J}\Big(\partial_{\varphi\varphi}\hat{u}_r(\rho,\varphi,\omega),\varphi_-\Big)=-\hat{f}(\omega-\omega_0)(i\omega\rho) e^{i\omega\rho}$.
From (\ref{^u_d}), (\ref{varepsilon_{pm}}), (\ref{maathcal{Z}}) and (\ref{8.0}) we have
\begin{equation}\label{mathcal J}
\mathcal{J}\Big(\hat{u}_d(\rho,\varphi,\omega),\varphi_-\Big)=\frac{i}{8\pi}\hat{f}(\omega-\omega_0)\int_{-1}^{1}\frac{4}{\beta+i\varepsilon} e^{i\omega\rho\cosh\beta}\;d\beta\Bigg\vert_{\varepsilon_-=+0}^{\varepsilon_{-}=-0}=-\mathcal{J}\Big(\hat{u}_r(\rho,\varphi,\omega),\varphi_-\Big).
\end{equation}
Further, by (\ref{dfz}) $\mathcal{J}\Big(\partial_{\varphi}\hat{u}_d(\rho,\varphi,\omega),\varphi_-\Big)=0=-\mathcal{J}\Big(\partial_{\varphi}\hat{u}_r(\rho,\varphi,\omega),\varphi_-\Big)$.\\
Finally, consider $M:=\mathcal{J}\Big(\partial_{\varphi\varphi}\hat{u}_d(\rho,\varphi,\omega),\varphi_-\Big)$. Similarly to (\ref{mathcal J}), expanding $e^{i\omega\rho\cos\beta}$ in the Taylor series in $\beta$ (in 0) and noting that all the terms $\ds\int\ds\frac{\beta^k\;d\beta}{(\beta+i\varepsilon_{-})^3},  k\neq2$ have jumps equal to 0, we obtain
\begin{equation*}
M=-\ds\frac{i}{\pi}\hat{f}(\omega-\omega_0)\int\limits_{-1}^{1}\frac{e^{i\omega\rho\cos\beta}}{(\beta+i\varepsilon_-)^3}d\beta\Bigg\vert_{\varepsilon_-=+0}^{\varepsilon_-=-0}=\ds\frac{\ds- i \hat{f}(\omega-\omega_0)(i\omega\rho )e^{i\omega\rho}}{2\pi}\int\limits_{-1}^{1}\frac{\beta^2}{(\beta+i\varepsilon_-)^3}d\beta\Bigg\vert_{\varepsilon_-=+0}^{\varepsilon_-=-0}.
\end{equation*}
Hence,
$
M=\hat{f}(\omega-\omega_0)(i\omega\rho )e^{i\omega\rho}=-\mathcal{J}\Big(\hat{u}_r(\rho,\varphi,\omega),\varphi_-\Big).
$
Since $\hat{u}_i^i(\rho,\varphi,\omega)$ is smooth on $l_{-}$ by (\ref{dil}), we obtain from (\ref{u^_s}) that $\hat{u}_{s}\in C^2(l_{-})$.

\no Similarly using (\ref{^u_r}), (\ref{u_1^1}) and (\ref{u_i}) we obtain: $\hat{u}_{s}\in C^2(l_{+})$. So $\hat{u}_{s}\in C^2(Q)$.  Proposition \ref{patial_varphi} is proven. ~~$\blacksquare$

\begin{coro}\label{6.7}
{\bf i)} The function $\hat{u}_s(\cdot,\cdot,\omega)$ belongs to the space $H^{1}(Q)$ for any $\omega\in\C^+$.

\no {\bf ii)} The function $u_{s}(x,t)\in \mathcal{M}$.

\end{coro}
\no {\bf Proof.} {\bf i)} Everywhere below $x=(\rho,\varphi)\in Q\setminus(l_1\cup l_2)$.
It suffices to prove that
\begin{equation}\label{bL_2}
u_{s}(\cdot,\cdot,\omega),~ \partial_{x_k}u_{s}(\cdot,\cdot,\omega)\in L_2(Q),\quad k=1,2,\quad \omega\in\C^{+}.
\end{equation}
First, by Proposition \ref{patial_varphi}, $\hat{u}_s(x,\omega)$ satisfies (\ref{5.6'}). Hence, $\hat{u}_s(\cdot,\omega)\in L_2(Q)$ for any $\omega\in\C^{+}$.\\
Further, using (\ref{x=}), we have $|\partial_{x_1}u_s(\cdot,\cdot,\omega)|^2\leq |\cos\varphi|^2 |\partial_{\rho}u_s(\cdot,\cdot,\omega)|^2+\frac{|\sin\varphi|^2}{\rho^2}|\partial_{\varphi}u_s(\cdot,\cdot,\omega)|^2$.
Hence, by Proposition \ref{patial_varphi} $|\partial_{x_1}u_s(\cdot,\cdot,\omega)|^2 \leq C(\omega) e^{-2c(\omega)}\Big(1+\ds\frac{1}{\rho}\Big)$.\\
This implies that $\partial_{x_1} u_s\in L_2(Q)$, since $c(\omega)>0$. Similarly, $\partial_{x_2}u_{s}(\cdot,\cdot,\omega)\in L_2(Q)$. (\ref{bL_2}) is proven.\\
{\bf ii)} The statement follows from Definition \ref{d1}.  ~~$\blacksquare$

\section{Uniqueness.}
\setcounter{equation}{0}

In Section\;5 we  proved the existence of solution to (\ref{D1})-(\ref{bc}) belonging to $\mathcal{M}$. In this section we prove the uniqueness of this solution in the same space.

\no Recall that we understand the uniqueness of the time-dependent Sommerfeld problem (\ref{mp})-(\ref{ic}) as the uniqueness of the solution $u_{s}$ given by (\ref{u_s'}) of the mixed problem (\ref{D1})-(\ref{bc}) in the space $\mathcal{M}$.
\begin{teo}\label{u}
The problem (\ref{D1})-(\ref{bc}) admits a unique solution in the space $\mathcal{M}$.
\end{teo}
\no {\bf Proof.}  We follow closely the proof of Theorem\;2.1 from \cite{g4}, exept that the angle in \cite{g4} can be now $2\pi$. Suppose that there exist two solutions $u_s(x,t)$ and $v_s(x,t)$ of system (\ref{D1})-(\ref{bc}) belonging to $\mathcal{M}$. Consider $w_s(x,t):=u_s(x,t)-v_s(x,t)$.\\
Then $\hat{w}_{s}(\cdot, \cdot,\omega)=\hat{u}_{s}(\cdot, \cdot,\omega)-\hat{v}_{s}(\cdot, \cdot,\omega)$, where $\hat{u}_s$, $\hat{v}_s$ and, therefore $\hat{w}_{s}$  satisfy all the conditions of Proposition \ref{patial_varphi} and $\hat{w}_{s}\vert_{W^0}=0$ by (\ref{BVP}).\\
Let us prove that $\hat{w}_s(\cdot, \cdot,\omega)\equiv 0$. Let $R$ be sufficiently large positive number and $B(R)$ be the open disk centred at the origin with radius $R$. Set $Q_{R}:=Q\cap B(R)$. Note that $Q_{R}$ has a piecewise smooth boundary $S_{R}$ and denote $n(x)$ the outward unit normal vector at the non-singular points $x\in S_{R}$   (see  Fig.\;{\;\re{Figg}}). \\
The first Green identity for $w_{s}(\rho,\varphi,\cdot)$ and its complex conjugate $\overline{w}_{s}$ in the domain $Q_{R}$, together with zero boundary conditions on $S_{R}$ yields
\begin{equation*}
\int_{Q_{R}} \Big[|\nabla\hat{w}_s|^2-\omega^2|\hat{w}_s|^2\Big]\;dx=\int\limits_{\partial B(R)\cap Q}\Big(\partial_{n} \hat{w}_s\Big)\cdot\Big(\overline{w}_{s}\Big)\;dS_{R}.
\end{equation*}
\newpage
\begin{figure}[htbp]
\centering
\includegraphics[scale=0.3]{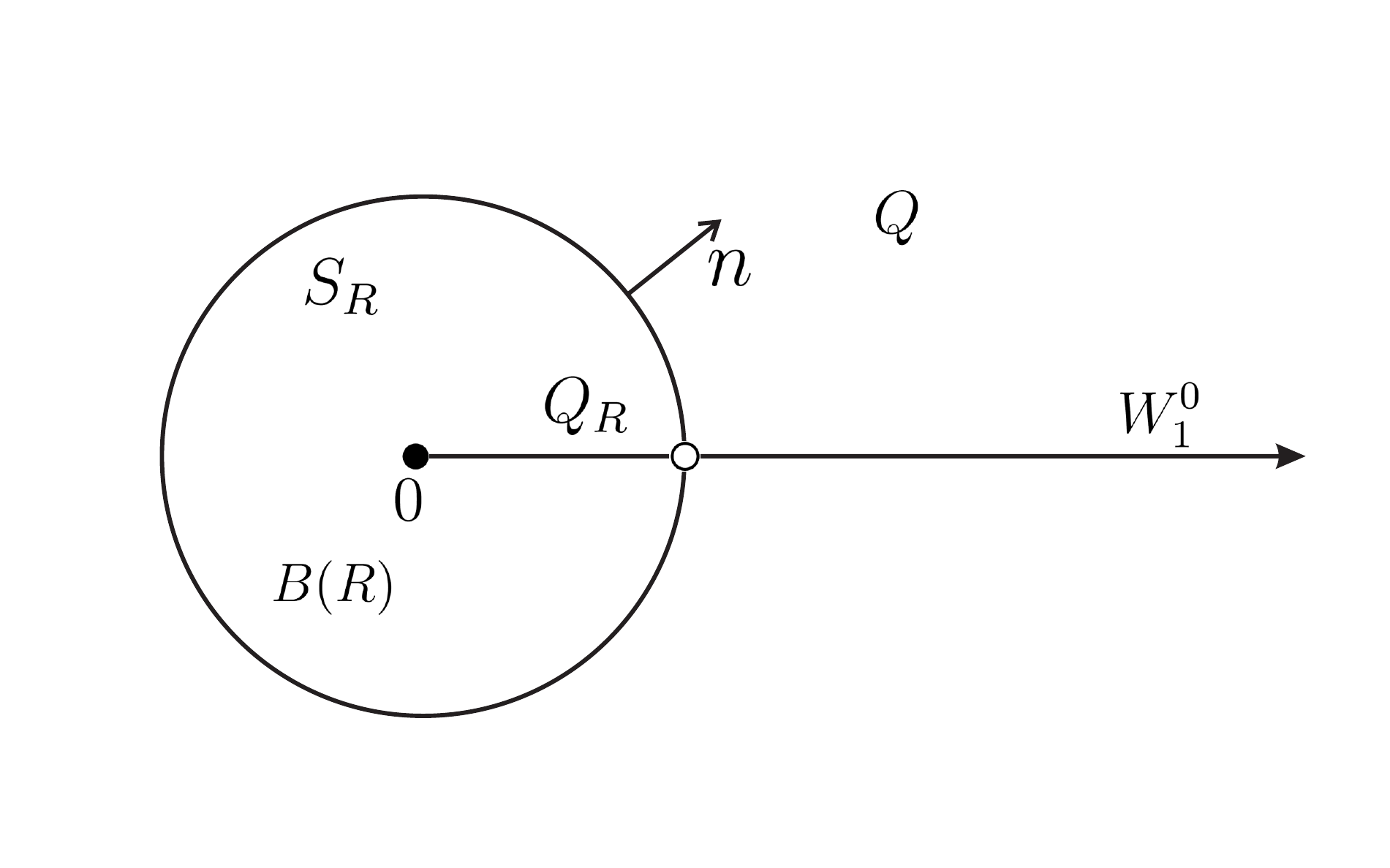}
\caption{Uniqueness}\label{Figg}
\end{figure}
\vspace{0.5cm}

\no From the real and imaginary parts of the last identity, we obtain
\begin{equation}\label{2.11}
\int\limits_{Q_{R}}\Big[|\nabla\hat{w}_{s}|^2+\big(\Im\omega\big)^2|\hat{w}_{s}|^2\Big]\;dx=\Re\int\limits_{\partial B(R)\cap Q}\big(\partial_{n}\hat{w}_s\big)\big(\overline{\hat{w}_{s}}\big)\;dS_{R}
\end{equation}
for $\Re\omega=0$ and
\begin{equation}\label{2.12}
-2\big(\Re\omega\big)\big(\Im\omega\big)\int\limits_{Q_{R}}|\hat{w}_s|^2\;dx=\Im\int\limits_{\partial B(R)\cap Q}\big(\partial_n w_s\big) \big(\overline{w}_s\big)\;dS_{R}
\end{equation}
for ${\rm Re}~\omega\neq0$.
Recall that we consider the case ${\rm Im}~k\neq0$. Now, note that since $\hat{\omega}_s\in H^1(Q)$, there exist a monotonic sequence of positive numbers $\{R_j\}$ such that $R_j\to\infty$ as $j\to\infty$ and
\begin{equation}
\label{2.13}
\lim\limits_{j\to\infty}\int\limits_{\partial B(R_j)\cap Q}\Big[\partial_n\hat{w}_s\Big]\Big[\overline{\hat{w}_s}\Big]dS_{R_j}=0.
\end{equation}
Indeed, in $(\rho,\varphi)$ polar coordinates, we have that the integrals
\begin{equation*}
\int\limits_0^\infty\left(R\int\limits_0^{2\pi}|\hat{w}_s(\rho,\varphi)|^2d\varphi\right)dR\quad{\rm and}\quad
\int\limits_0^\infty\left(R\int\limits_0^{2\pi}|\partial_n\hat{w}_s(\rho,\varphi)|^2d\varphi\right)dR
\end{equation*}
are finite. This fact, in particular, implies that there exist a monotonic sequence of positive numbers $R_j$ such that $R_j\to\infty$ as $j\to\infty$ and
\begin{equation*}
\int\limits_0^{2\pi}|\hat{w}_s(R_j,\varphi)|^2d\varphi=o(R_j^{-1})\quad{\rm and}\quad
\int\limits_0^{2\pi}|\partial_n\hat{w}_s(R_j,\varphi)|^2d\varphi=o(R_j^{-1}),\quad{\rm as}\quad j\to\infty.
\end{equation*}
Further, applying the Cauchy-Schawrtz inequality for every $R_j$, we get
\begin{equation*}
\begin{array}{llll}
\ds\left|\int\limits_0^{2\pi}\partial_n\hat{w}_s(R_i,\varphi)\overline{\hat{w}_s}(R_i,\varphi)d\varphi\right|&\leq&\ds
\int\limits_0^{2\pi}|\partial_n\hat{w}_s(R_i,\varphi)\hat{w}_s(R_i,\varphi)|d\varphi\leq\\
\\
&\leq&\ds\left(\int\limits_0^{2\pi}|\partial_n\hat{w}_s(R_i,\varphi)|^2d\varphi\right)^{1/2}
\left(\int\limits_0^{2\pi}|\hat{w}_s(R_i,\varphi)|^2d\varphi\right)^{1/2}\\
\\
&=&o(R^{-1}_j)\quad{\rm as}\quad j\to\infty
\end{array}
\end{equation*}
and  therefore we obtain (\ref{2.13}).\\
Since the expressions under the integral sign in the left hand side of the equalities (\ref{2.11}) and (\ref{2.12}) are non-negative, then we have that these integrals are monotonic with respect to $R$. This observation together with (\ref{2.13}) implies
\begin{equation*}
\int\limits_{Q}\Big[|\nabla\hat{w}_{s}|^2+\big(\Im\omega\big)^2|\hat{w}_{s}|^2\Big]\;d\varphi=\lim\limits_{R\to\infty}
\int\limits_{Q_R}\Big[|\nabla\hat{w}_{s}|^2+\big(\Im\omega\big)^2|\hat{w}_{s}|^2\Big]\;d\varphi=0
\end{equation*}
for ${\rm Re}~\omega=0$ and $\ds\int\limits_{Q}|\hat{w}_{s}|^2\;d\varphi=\lim\limits_{R\to\infty}
\int\limits_{Q_R}|\hat{w}_{s}|^2\;d\varphi=0$
for ${\rm Re}~\omega\neq0$.
Thus, it follows from the last two identities that $\hat{w}_{s}=0$ in $Q$. ~~~$\blacksquare$

\section{Conclusion}

We  proved that the Sommerfeld solution to the half-plane diffraction problem for a wide class of incident waves is the limiting amplitude of the solution of the corresponding time-dependent problem in a functional class of generalized solutions. The solution of the time-dependent problem is shown to be unique in this class. It is also shown that the limiting amplitude automatically satisfies the Sommerfeld radiation condition and the regularity edge condition.

\section{Appendix I.}
\setcounter{equation}{0}

\begin{lem}\label{Z^0(z_)}
i) The functions $\mathcal{Z}$ (given by (\ref{CZ})) and $\partial_\varphi \mathcal{Z}$ admit  uniform with respect to $\varphi\in [0,2\pi]$ estimates
\begin{equation}\label{asH^0}
|\mathcal{Z}(\beta,\varphi)|\leq C e^{-|\beta|/2},\quad |\partial_\varphi\mathcal{Z}(\beta,\varphi)|\leq C e^{-|\beta|/2},\quad
|\beta|\geq1.
\end{equation}
ii) The function $\mathcal{Z}$ admits the representation
\begin{equation}\label{maathcal{Z}}
\mathcal{Z}(\beta,\varphi)=-\frac{4}{\beta+i\varepsilon_+}+\frac{4}{\beta+i\varepsilon_-}+\mathcal{\check{Z}}(\beta,\varphi),\quad \varepsilon_{\pm}\neq0
\end{equation}
with
\begin{equation}\label{8.0}
 \mathcal{\check{Z}}(\beta,\varphi)\in C^{\infty}(\R\times[0,2\pi]),\quad |\mathcal{\check{Z}}(\beta,\varphi)|\leq C,\quad \beta\in\R\times[0,2\pi].
\end{equation}
\end{lem}
 \no iii) $\partial_\varphi\mathcal{Z}$ admits the representation
 \begin{equation}
\label{dfz}
\partial_\varphi\mathcal{Z}=-\frac{4i}{(\beta+i\varepsilon_+)^2}+\frac{4i}{(\beta+i\varepsilon_-)^2}+\check{Z}_1(\beta,\varphi), \quad\varepsilon_{\pm}\neq0,
\end{equation}
with
\begin{equation}
\label{ez1}
 \quad
\check{Z}_1(\beta,\varphi)\in C^\infty(\R\times[0,2\pi]),\quad|\check{Z}_1(\beta,\varphi)|\leq C,\quad \beta\in\R\times[0,2\pi].
\end{equation}
{\bf Proof.} i) For $a=im, b=in$, we have $\coth a-\coth b=\ds\frac{-\sinh(\alpha/2)}{\sinh(b)\sinh (a)}$.
Hence for
$
m=-\pi/8+a/4, ~ n=-\pi/8-a/4
$
we obtain the estimate (\ref{asH^0}) for $U(\zeta)$ given by (\ref{H^{0}}) with respect to $\zeta$. So (\ref{asH^0}) for $\mathcal{Z}$ follows from (\ref{Z^{0}}), (\ref{CZ}).\\
ii) From (\ref{Z^{0}}), (\ref{H^{0}}) it follows that the function $\mathcal{Z}$  admits the representation
\begin{equation*}
\mathcal{Z}(\beta,\varphi)=Z_+(\beta,\varphi)+Z_-(\beta,\varphi)+Z^{+}(\beta,\varphi)+Z^{-}(\beta,\varphi),
\end{equation*}
where
\begin{equation}\label{Z_k}
\left.\ba{rcl}
Z_\pm(\beta,\varphi)=\pm \coth\Bigg(\ds\frac{\beta+i(\varphi_{\pm}-\varphi)}{4}\Bigg),\quad Z^{\pm}(\beta,\varphi)=\pm \coth\Bigg(\ds\frac{\beta-i\big(\varphi_{\pm}+\varphi\big)}{4}\Bigg).
\ea
\right.
\end{equation}
Further, since $|\coth z-1/z |\leq C$, $|{\rm Im~}z|\leq\pi$, $z\neq0$,
$Z_\pm(\beta,\varphi)=\ds \pm\frac{4}{\beta+i\varepsilon_\pm}+\check{Z}_\pm(\beta,\varphi),~ \varphi\neq\varphi_\pm$, $\check{Z}_\pm(\beta,\varphi)\in C^\infty(\R\times[0,2\pi]),~|\check{Z}_\pm(\beta,\varphi)|\leq C,~ \beta\in\R\times[0,2\pi]$.\\
Finally, by (\ref{alpha}), $Z^{\pm}(\beta,\varphi)\in C^\infty(\R\times[0,2\pi])$ and
$|Z^{\pm}(\beta,\varphi)|\leq C,~ \beta\in\R\times[0,2\pi]$.\\
Therefore, (\ref{maathcal{Z}}), (\ref{8.0}) are proven.\\
iii) From (\ref{maathcal{Z}}), (\ref{varepsilon_{pm}}) we get (\ref{dfz}). Finally, $\partial_\varphi Z^{\pm}(\beta,\varphi)\in C^\infty(\R\times[0,2\pi])$, and $|\partial_\varphi Z^{\pm}(\beta,\varphi)|\leq C,~(\beta,\varphi)\in\R\times[0,2\pi]$,  by (\ref{Z_k}). Moreover, since $\partial_\varphi Z_\pm(\beta,\varphi) \pm [4i/(\beta+\varepsilon_\pm)^2]\in C^\infty([\R\times[0,2\pi])$, and is bounded in the same region, (\ref{ez1}) holds.  ~~$\blacksquare$\\

\medskip

\no For $\varepsilon,\beta\in\R,~\varepsilon\neq0,~\rho>0,~\omega\in\C^{+}$, let
\begin{equation}\label{K}
K_0(\beta,\rho,\omega,\varepsilon):=\frac{e^{i\omega\rho\cosh\beta}}{\beta+i\varepsilon},\quad \mathcal{K}_0(\rho,\omega,\varepsilon):=\int\limits_{-1}^{1} K(\beta,\rho,\omega,\varepsilon)\;d\beta
\end{equation}
\begin{equation}\label{dk}
K_1(\beta,\rho,\omega,\varepsilon):=\cosh\beta\cdot e^{i\omega\rho\cosh\beta},\quad \mathcal{K}_1(\rho,\omega,\varepsilon)=\int\limits_{-1}^{1}K_1(\beta,\rho,\omega,\varepsilon)\;d\beta
\end{equation}
\begin{equation*}
K_2(\beta,\rho,\varphi,\varepsilon):=\frac{e^{i\omega\rho\cosh\beta}}{(\beta+i\varepsilon)^2},\quad \mathcal{K}_2(\rho,\omega,\varepsilon):=\int\limits_{-1}^1 K_2(\beta,\rho,\omega,\varepsilon)\;d\beta\;d\beta.
\end{equation*}
\begin{lem}\label{lcI}
There exist $C(\omega)>0$, $c(\omega)>0$ such that the functions $\mathcal{K}_0, \mathcal{K}_1, \mathcal{K}_2$ satisfy the estimates
\begin{equation}\label{K012}
|\mathcal{K}_{0,1,2}(\rho, \omega, \varepsilon)|\leq C(\omega)e^{-c(\omega)\rho},\quad \rho>0,~\varphi\in(0,2\pi),~\varepsilon\neq0.
\end{equation}

\end{lem}
\no{\bf Proof.} It suffices to prove (\ref{K012}) for $0<\varepsilon<\varepsilon_0$, since the functions $\mathcal{K}_0, \mathcal{K}_1, \mathcal{K}_2$ are odd with respect to $\varepsilon$, and for $\varepsilon\geq\varepsilon_0>0$ they satisfy the estimate
\begin{equation*}
\begin{array}{rl}
\Big|\ds\mathcal{K}_{0,1,2}(\beta,\rho,\omega,\varepsilon)\Big|\leq C(\varepsilon_0) \ds\int\limits_{-1}^{1} e^{-\omega_2\rho}\;d\beta\leq 2C(\varepsilon_0) e^{-\omega_2\rho}.
\end{array}
\end{equation*}
I) Let us prove (\ref{K012}) for $\mathcal{K}_0$. Let
\begin{equation}\label{ch}
\cosh\beta:=1+h(\beta),\quad\beta\in\C.
\end{equation}
Define
$\varepsilon_0=\varepsilon_0(\omega)$ such that
\begin{equation}\label{h}
|h(\beta)|<\frac{1}{4},\quad |\omega_1||h(\beta)|\leq \frac{\omega_2}{4},\quad ~{\rm for}~ |\beta|\leq2\varepsilon_0:=r,
\end{equation}
define the contour
\begin{equation}\label{gamma_r}
\gamma_{r}:=\lbrace \beta=r e^{i\theta},\quad -\pi<\theta<0\rbrace.
\end{equation}
Then we have by the Cauchy Theorem $\mathcal{K}_0(\rho,\omega,\varepsilon)=I_1(\rho,\omega,\varepsilon)+I_2(\rho,\omega,\varepsilon)-2\pi i\;
\Res_{\beta=-i\varepsilon} K_0(\beta,\rho,\omega,\varepsilon)$, where $I_1(\rho,\omega,\varepsilon)=\int\limits_{\gamma_r} K_0(\beta,\rho,\omega,\varepsilon)\;d\beta, \quad
I_2(\rho,\omega,\varepsilon)=\Bigg(\ds\int\limits_{-1}^{-r}+\int\limits_{r}^{1}\Bigg) K_0(\beta,\rho,\omega,\varepsilon)\;d\beta,~ 0<\varepsilon<\varepsilon_0$.\\
First,
\begin{equation}\label{res}
|\Res_{\beta=-i\varepsilon} K_0(\beta,\rho,\omega,\varepsilon)|=e^{-\omega_2\rho\cos\varepsilon}\leq e^{-\frac{1}{2}\omega_2\rho},\quad 0<\varepsilon<\varepsilon_0,
\end{equation}
 by (\ref{h}). Further, from  (\ref{ch}) we have
\begin{equation}\label{esI}
\big|I_1(\rho,\omega,\varepsilon)\big|\leq \int\limits_{\gamma_r}\frac{\Big|e^{-\omega_2\rho\big(1+h(\beta)\big)} e^{i\omega_{1}\rho\big(1+h(\beta)\big)}\Big|}{|\beta+i\varepsilon|}|d\beta|\leq \frac{1}{\varepsilon_0}\;e^{-\omega_2\rho}\int\limits_{\gamma_r} \big|e^{-\omega_2\rho\;h(\beta)+i\omega_1\rho\;h(\beta)}\big|\;\big|d\beta\big|,
\end{equation}
since for $\beta\in\gamma_{r}$ we have $|\beta+i\varepsilon|\geq |\beta|-\varepsilon=2\varepsilon_0-\varepsilon>\varepsilon_0$, see Fig. \re{Fige}.

\begin{figure}[htbp]
\centering
\includegraphics[scale=0.3]{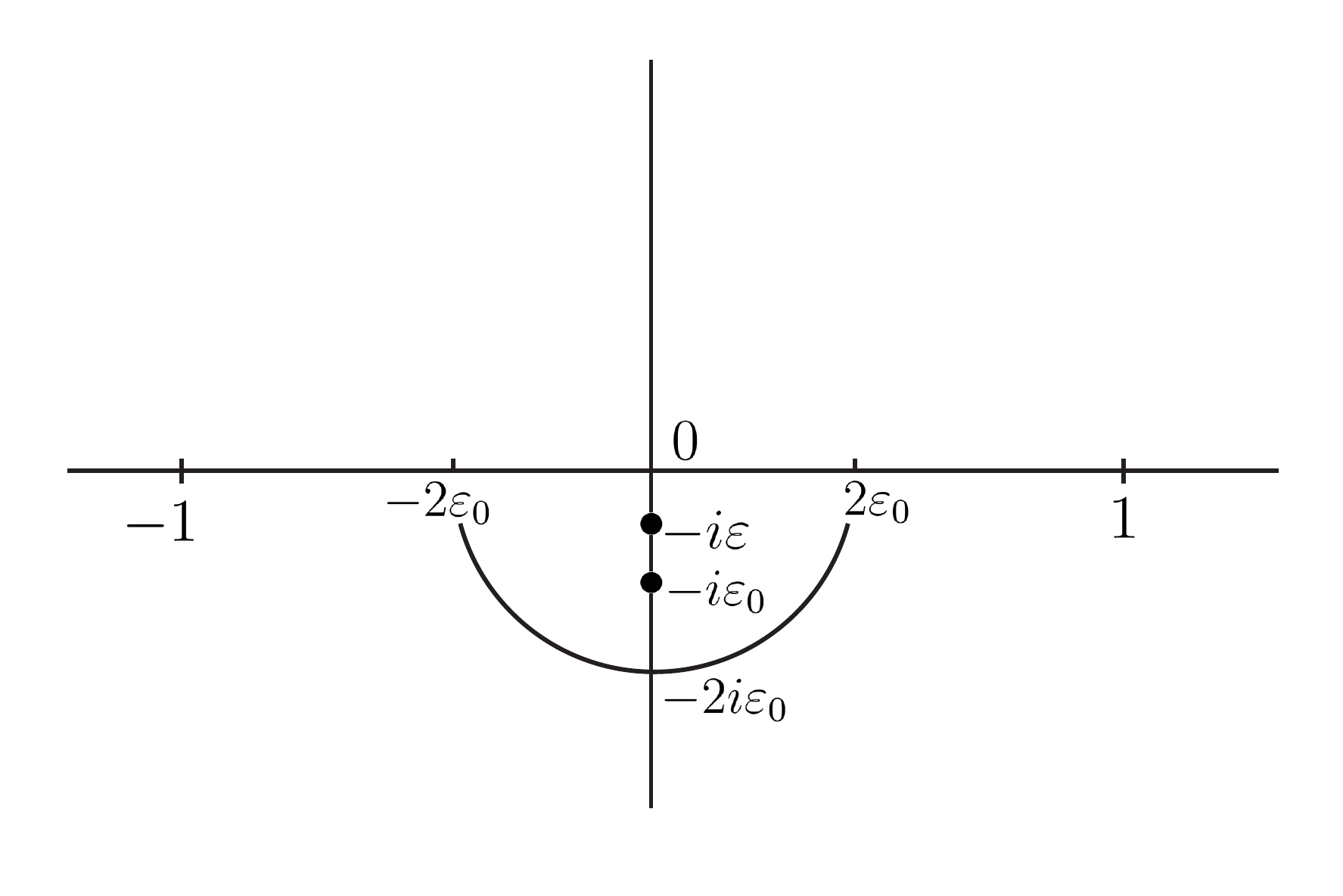}
\caption{Contour $\gamma_r$}\label{Fige}
\end{figure}


\no Let $h(\beta):=h_1(\beta)+ih_2(\beta)$. Then
\begin{equation}\label{I_1}
|I_1(\rho,\omega,\varepsilon)|\leq\ds\frac{1}{\varepsilon_0}\;e^{-\omega_2\rho}\ds\int_{\gamma_r} e^{\omega_2\rho\;|h_1(\beta)|}\;e^{|\omega_1|\rho\;|h_2(\beta)|}\;d\beta\leq 2\pi e^{-\omega_2\rho/2},
\end{equation}
by (\ref{h}). Finally,
\begin{equation}\label{I_2}
|I_2(\rho,\omega,\varepsilon)|\leq\ds\int_{[-1,-r]\cup[r,1]}\Bigg|\ds\frac{e^{-\omega_2\rho\cosh\beta+i\omega_1\rho\cosh\beta}}{\beta+i\varepsilon}\Bigg|\;d\beta
\leq\ds\frac{1}{2\varepsilon_0(\omega)}\;e^{-\omega_2\rho},
\end{equation}
since $|\beta+i\varepsilon|\geq2\varepsilon_0,~\beta\in[-1,-r]\cup[r,1]$. From (\ref{esI})-(\ref{I_2}), we obtain (\ref{K012}) for
$\mathcal{K}_0$.\\
II) Let us prove (\ref{K012}) for $\mathcal{K}_1$.  Let $h(\beta), \varepsilon_0(\omega), \gamma_{r}$ be defined by (\ref{ch})-(\ref{gamma_r}) hold. Then we have by the Cauchy Theorem
\begin{equation}\label{4.65'}
\mathcal{K}_1(\rho,\omega,\varepsilon):= \int\limits_{\gamma_r\cup[-1,r]\cup[r,1]} K_1(\beta,\rho,\omega,\varepsilon)\;d\beta-2\pi i\;\Res_{\beta=-i\varepsilon}\;K_1(\beta,\rho,\omega,\varepsilon),\quad 0<\varepsilon<\varepsilon_0.
\end{equation}
First, similarly to (\ref{res}), we obtain $\big|\Res_{\beta=-i\varepsilon}\;K_1(\beta,\rho,\omega,\varepsilon)\big|\leq |\omega| e^{-\frac{\omega_2\rho}{2}}$,
by (\ref{h}). Further, by (\ref{h}) similarly to the proof of (\ref{esI}),(\ref{I_1}), and using (\ref{ch}), we get
\begin{equation}\label{4.18'}
\ds\Bigg|\int\limits_{\gamma_r} K_1(\beta,\rho,\omega,\varepsilon)\;d\beta\Bigg|
\leq \frac{|\omega|}{\varepsilon_0}\cdot\frac{5}{4}\; e^{-\omega_2\rho}\int\limits_{\gamma_r} |e^{-\omega_2\rho\; h(\beta)}\;e^{i\omega_1\rho\; h(\beta)}|\;|d\beta|\leq C(\omega) e^{-\frac{\omega_2\rho}{2}}
\end{equation}
Finally, similarly to the proof of (\ref{I_2}) we get the estimate
\begin{equation}\label{4.67'}
\Bigg|\ds\int\limits_{[-1,-r]\cup[r,1]} K_1(\beta,\rho,\omega,\varepsilon)\;d\beta\Bigg|\leq C(\omega)\;e^{-\omega_2\rho}.
\end{equation}
From (\ref{4.65'})-(\ref{4.67'}), we obtain (\ref{K012}) for $\mathcal{K}_1$.

\medskip
\no III) Estimate (\ref{K012}) for $\mathcal{K}_2$ is proved similarly to the same estimate for $\mathcal{K}_{0,1}$ with the obvious changes.
Lemma \ref{lcI} is proven. ~~$\blacksquare$

\section{Appendix II.}
\setcounter{equation}{0}
\begin{lem}\label{lu_d}
\begin{equation}\label{chu_d}
\Big(\Delta+\omega^2\Big)u_d(\rho,\varphi,\omega)=0,\quad \varphi\neq\varphi_{\pm},\quad \omega\in\C^{+}.
\end{equation}
\end{lem}
{\bf Proof.} By (\ref{^u_d}) it suffices to prove (\ref{chu_d}) for
\begin{equation}\label{S_d}
A_d(\rho,\varphi,\om):=\int\limits_{\R} \mathcal{Z}(\beta,\varphi) e^{i\omega\rho\cosh\beta}\;d\beta.
\end{equation}
Since $\omega\in\C^{+}$ the integral (\ref{S_d}) converges after differentiation with respect to $\rho$ and $\varphi$. We have
\begin{equation*}
\partial_{\rho}A_{d}(\rho,\varphi,\omega)= (i\omega)\ds\int\limits_{\R} \mathcal{Z}(\beta,\varphi) \cosh\beta\;e^{i\omega\rho\cosh\beta}\;d\beta,\quad \partial^{2}_{\rho}A_{d}(\rho,\varphi,\omega)=-\omega^2\ds\int\limits_{\R} \mathcal{Z}(\beta,\varphi) \cosh^2\beta\; e^{i\omega\rho\cosh\beta}\;d\beta.
\end{equation*}
Integrating by parts,  we have by (\ref{CZ}) and (\ref{asH^0})
\begin{equation}\label{d+}
\begin{array}{lll}
\partial_{\varphi}A_{d}(\rho,\varphi,\omega)=\ds\int_{\R} \partial_{\varphi}\Big(Z^{0}(\beta+2\pi i-i\varphi) \Big) e^{i\omega\rho\cosh\beta}\;d\beta
 = -\omega\rho \ds\int\limits_{-\infty}^{\infty}\mathcal{Z}(\beta,\varphi)\sinh\beta\; e^{i\omega\rho\cosh\beta}\;d\beta,\quad \varphi\neq\varphi_{\pm}.
\end{array}
\end{equation}
 Hence, similarly to (\ref{d+}) $\partial^2_{\varphi\varphi}A_{d}(\rho,\varphi,\omega)= -i\omega\rho \ds\int\limits_{-\infty}^{\infty} \mathcal{Z}(\beta,\varphi)\Big[\cosh\beta+i\omega\rho\sinh^2\beta\Big]\;e^{i\omega\rho\cosh\beta}\;d\beta$, and
\begin{equation*}
\begin{array}{lll}
(\Delta+\omega^2) u_{d}(\rho,\varphi,\omega)
= \partial^2_{\rho} A_{d}(\rho,\varphi,\omega)+\ds\frac{1}{\rho}\partial_{\rho} A_{d}(\rho,\varphi,\omega)+\ds\frac{1}{\rho^2}\partial^2_{\varphi} A_{d}(\rho,\varphi,\omega)+\omega^2 A_{d}(\rho,\varphi,\omega)=0. ~~~\blacksquare
\end{array}
\end{equation*}

\section*{Acknowledgements}
A.E. Merzon and P. Zhevandrov are supported by CONACYT and CIC of UMSNH, M\'exico.\\
J.E. De la Paz M\'endez and T.J. Villalba Vega are supported by CONACYT, M\'exico.


\end{document}